\documentclass{nature}
\pdfoutput=1

\newcommand{\uc}{\overline{u}}
\newcommand{\wc}{\overline{w}}

\newcommand{\umx}{m_{\mathrm{x}}}
\newcommand{\umy}{m_{\mathrm{y}}}
\newcommand{\umz}{m_{\mathrm{z}}}

\usepackage{graphicx}
\usepackage{amssymb}
\usepackage{amsmath}
\usepackage{mcite}
\usepackage{xcolor}
\usepackage[normalem]{ulem}
\usepackage{bm}
\usepackage{color,soul}
\usepackage{gensymb}
\usepackage{ulem}
\newcommand{\onlinecite}[1]{\hspace{-1 ex} \nocite{#1}\citenum{#1}}
\renewcommand\vec{\mathbf}
\usepackage{lineno}
\makeatletter
\let\saved@includegraphics\includegraphics
\AtBeginDocument{\let\includegraphics\saved@includegraphics}
\renewenvironment*{figure}{\@float{figure}}{\end@float}
\makeatother

\usepackage{xcolor,soul}

\title{Experimental Observation of Vortex Rings in a Bulk Magnet}

\author{Claire Donnelly$^{1,2,3}$, Konstantin L. Metlov$^{4,5}$, Valerio Scagnoli$^{2,3}$, Manuel Guizar-Sicairos$^3$, Mirko Holler$^3$, Nicholas S. Bingham$^{2,3}$, J{\"o}rg Raabe$^3$, Laura J. Heyderman$^{2,3}$, Nigel Cooper$^{1}$ and Sebastian Gliga$^{3}$}

\begin{document}
\colorlet{hlcolorname}{yellow!30}
\sethlcolor{hlcolorname}

\maketitle

\begin{affiliations}
 \item Cavendish Laboratory, University of Cambridge, JJ Thomson Ave, Cambridge CB3 0HE, UK.
 \item Laboratory for Mesoscopic Systems, Department of Materials, ETH Z{\"u}rich, 8093 Z{\"u}rich, Switzerland.
 \item Paul Scherrer Institute, 5232 Villigen PSI, Switzerland.
 \item Donetsk Institute for Physics and Engineering, R. Luxembourg 72, Donetsk 83114, Ukraine.
 \item Institute for Numerical Mathematics RAS, 8~Gubkina str., 119991~Moscow GSP-1, Russia.
\end{affiliations}

\begin{abstract}

Vortex rings are remarkably stable structures occurring in { numerous systems:} for example in {turbulent} gases, where they are at the origin of weather phenomena\cite{Yao1996}; in fluids with implications for biology\cite{Kilner2000}; in electromagnetic discharges\cite{Stenhoff-book}; and in plasmas\cite{Akhmetov-book}. While vortex rings have also been predicted to exist in ferromagnets \cite{Cooper99}, they have not yet been observed. 
Using X-ray magnetic nanotomography\cite{donnelly17}, we imaged three-dimensional structures forming closed loops in a bulk micromagnet, { each composed of a vortex-antivortex pair}. Based on the magnetic vorticity, a quantity analogous to hydrodynamic vorticity, we identify these configurations as magnetic vortex rings. 
While such structures have been predicted to exist  as transient states in exchange ferromagnets\cite{Cooper99}, the vortex rings we observe exist as stable, static configurations, whose stability we attribute to the dipolar interaction. In addition, we observe stable vortex loops intersected by magnetic singularities \cite{Feldtkeller1965}, at which  the magnetisation within the vortex and antivortex cores reverses. We gain insight into the stability of these states through field and thermal equilibration protocols. These measurements pave the way for the observation of complex three-dimensional solitons in bulk magnets, as well as for the development of  applications based on three-dimensional magnetic structures.
\end{abstract}

In magnetic thin films, vortices are naturally occurring flux closure states, in which the magnetisation curls around a stable core, where the magnetisation tilts out of the film plane~\cite{SOHSO00,WWBPMW02}. 
These structures have been studied extensively over the past decades due to their intrinsic stability~\cite{G2008} and their topology-driven dynamics~\cite{Choe2004,Waeyenberge2006,HGFS07}, which are of both fundamental and technological~\cite{Pigeau2010} interest.
Antivortices, the topological counterpart of vortices, distinguish themselves from vortices by an opposite rotation of the in-plane magnetization that is quantified by the index of the vector field -- which is equal to the winding number of a path traced by the magnetisation vector while moving in the  counterclockwise direction around the core~\cite{HS2006}. While vortices have a circular symmetry of the magnetisation (figure 1a), antivortices only display inversion symmetry about the center~\cite{gliga2008} (figure 1b), resembling saddle points in the vector field. 
Experimental studies of magnetic vortices and antivortices have mostly been restricted to two dimensional, planar systems, in which vortex-antivortex pairs have a natural tendency to annihilate~\cite{gligaJAP2008}, unless they are part of larger, stable structures, such as cross-tie walls~\cite{Neudert2006}. 

\begin{figure}
\centering
\includegraphics[width=0.9\columnwidth]{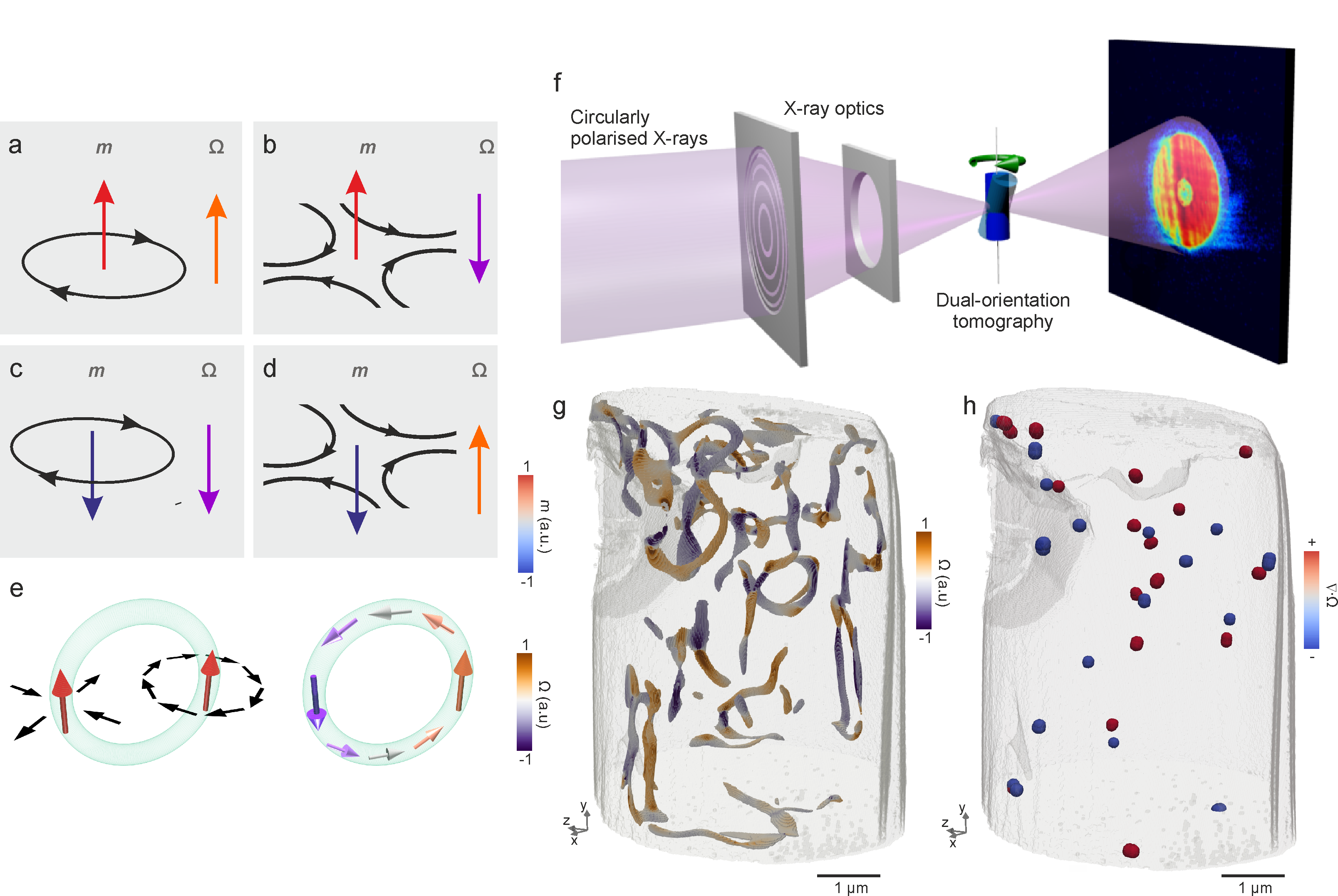}
\caption{\label{fig:fig1} Measuring and reconstructing the internal magnetic structure and the magnetic vorticity within a GdCo$_2$ pillar. a-d) Schematic representation of the magnetic vorticity $\vec{\Omega}$, shown in purple and orange arrows, for a number of vortex and antivortex configurations with different polarisations (red, blue). The vorticity of a vortex-antivortex pair with same polarisation is shown in (e). f) Schematic representation of the experimental setup: tomographic projections with magnetic contrast are measured using dichroic ptychography for the sample at several different orientations with respect to the X-ray beam. Measurements were performed with the sample at two different tilt angles: 30\degree~(transparent green) and 0\degree~(blue).  g) Plotting regions of significant magnetic vorticity, we locate a network of structures, and h) plotting regions of high divergence of the vorticity $\vec{\nabla}\cdot\vec{\Omega}$, we locate Bloch points (red) and anti-Bloch points (blue), which respectively have positive and negative divergence.}
\end{figure}

In bulk ferromagnets, the existence of transient vortex rings, that take the form of localised  solitons and are analogous to smoke rings, has been predicted~\cite{Cooper99}, but such structures have so far not been observed. Just as vortex rings in fluids are characterised by their vorticity, ferromagnetic vortex ring structures can be identified by considering the magnetic vorticity.  By analogy with fluid vorticity, the magnetic vorticity is a vector field, which can be defined as~\cite{Cooper99}:
\begin{equation}\label{eq:magvort}
    \Omega_\alpha = \frac{1}{8\pi}\epsilon_{\alpha\beta\gamma}\epsilon_{ijk}n_i\partial_\beta n_j \partial_\gamma n_k
\end{equation}
where $n_\alpha(\mathbf{r},t)$ is a component of the unit vector representing the local orientation of the magnetisation, $\alpha$ indicates the vorticity component, and { $\epsilon_{\alpha\beta\gamma}$ is the Levi-Civita tensor, summed over three components $x,y,z$.}
The magnetic vorticity vector $\mathbf{\Omega}$ represents the topological charge flux\cite{BP75} (or Skyrmion number\cite{Senthil2004}) density. Integrating the magnetic vorticity over a closed two-dimensional surface, results in a scalar value $\int \mathbf{\Omega}\cdot\mathrm{d}S=N$ corresponding to the skyrmion number, which gives the degree of mapping of the magnetization distribution to an order parameter space described by the surface of an $S^2$ sphere.
When  $N=1$, the target sphere is wrapped exactly once and each direction of the magnetisation vector is present.
The magnetic vorticity vector $\mathbf{\Omega}$ is therefore non-vanishing in the vicinity of the cores of vortices or antivortices, and is  represented in Figure \ref{fig:fig1}a-d for vortices and antivortices with different polarisations (the polarisation is the orientation of the magnetisation within the core). The vorticity vector is aligned parallel to the polarisation of a vortex (a,c) and antiparallel to the polarisation of an antivortex (b,d), indicating that it is dependent upon the direction of the magnetisation in the core as well as the index of the structure. Consequently, a vortex-antivortex pair with parallel polarisations, exhibit opposite vorticities, that circulate in a closed loop (Figure ~\ref{fig:fig1}e).

Here, we use the magnetic vorticity to locate and identify magnetic structures within a three-dimensional magnetic micropillar, that are imaged using hard X-ray magnetic nanotomography.  Within the bulk of the pillar, we find two types of vorticity loops. The first is characterised by a circulating magnetic vorticity forming vortex rings, analogous to smoke rings. These magnetic vortex rings consist of vortex-antivortex pairs with parallel polarisations, as in Figure~\ref{fig:fig1}e. Consequently, the magnetisation distribution does not wrap the order parameter space and the pair belongs to the same topological sector as a uniformly magnetised domain. The second type of loop contains singularities, or Bloch points\cite{Feldtkeller1965}, at which the vorticity abruptly reverses its sign, thus modifying the topology of the vortex-antivortex pairs. Calculating preimages of the observed structures indicates that the vortex rings display concentric pre-images corresponding to a trivial knot, with a vanishing Hopf index, while structures containing Bloch points have preimages similar to recently observed `toron' structures in anisotropic fluids~\cite{Ackerman2017}.

The hard X-ray magnetic nanotomography setup is illustrated in Figure~\ref{fig:fig1}f. During the measurement, high resolution X-ray projections of the sample were measured with dichroic ptychography\cite{Donnelly2016} for 1024 orientations of the sample with respect to the X-ray beam. The photon energy of the circularly-polarised X-rays was tuned to the Gd $L_3$ edge and, by exploiting the X-ray magnetic circular dichroism effect, sensitivity to the component of the magnetisation parallel to the X-ray beam was obtained. In order to gain access to all three components of the magnetisation, X-ray projections were measured for different sample orientations about the tomographic rotation axis for two different sample tilts.  The internal magnetic structure was obtained using an iterative reconstruction algorithm\cite{donnelly17}, which has been demonstrated to offer a robust reconstruction of nanoscale magnetic textures\cite{Donnelly2018}. Further experimental details are given in the Methods section.

Using this method, we image the magnetic structure of a bulk GdCo$_\mathrm{2}$ ferrimagnetic cylinder of diameter 5 $\mathrm{\mu m}$, in which the coupling between the two antiparallel magnetic sublattices leads to an effective soft ferromagnetic behavior\cite{chikazumi-book}. The lowest energy state of such a magnetic cylinder { is expected to} consist of a single vortex\cite{AHA79}. In our system, the size of the pillar is large enough to reduce the role of surface anisotropy, { supporting the stabilisation of more complex, metastable }states, that can include a large number of vortices, anti-vortices, domain walls and singularities\cite{donnelly17}. 

We compute the magnetic vorticity $\mathbf{\Omega}$  from the reconstructed magnetisation following equation (\ref{eq:magvort}). 
Regions of large vorticity are plotted in  Figure \ref{fig:fig1}g, where a number of `tubes' and loops corresponding to the cores of vortices and antivortices are visible. In addition, unlike in incompressible fluids, where the divergence must vanish, a non-zero divergence of the magnetisation, $\mathbf{m}$, is allowed in ferromagnets, given that Maxwell's equations only exclude the divergence of $\mathbf{B}$.
In this way, computing the magnetic vorticity also allows us to locate singularities of the magnetisation -- known as Bloch points -- within the system, which are characterised by a very large divergence of the magnetic vorticity, $\vec{\nabla}\cdot\vec{\Omega}$, due to the local variation in the orientation of the magnetisation.  Here, Bloch point and anti-Bloch points are identified by positive (red) and negative (blue) $\vec{\nabla}\cdot\vec{\Omega}$, as plotted in Figure \ref{fig:fig1}h.   Within the pillar, we find an equal number of Bloch points and anti-Bloch points, indicating that the singularities originated in the bulk of the structure, where they can only be created in pairs. As a result, it appears that sample boundaries, through which a single Bloch point could be injected, most likely did not play a role in the formation of the observed structures.

\begin{figure}
\centering
\includegraphics[width=0.9\columnwidth]{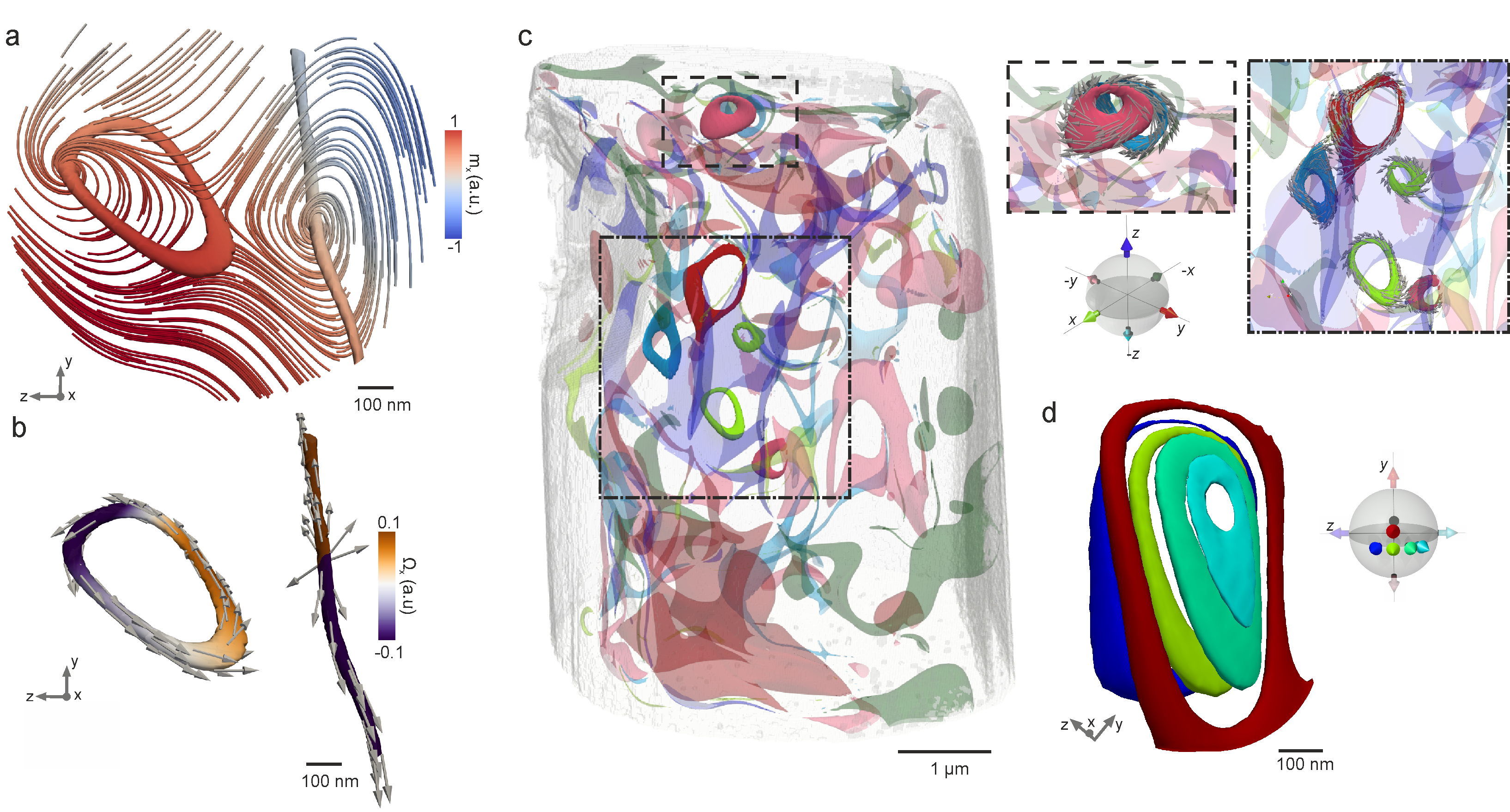}
\caption{\label{fig:fig3} 
Structure of a  vortex ring with circulating magnetic vorticity.
a) A vorticity  `loop'  is identified next to a vortex by plotting an isosurface corresponding to $m_x=\pm1$. The in-plane magnetisation within a two-dimensional slice through the loop is plotted using streamlines, revealing  two vortices enclosing an antivortex, with the loop consisting of a vortex-antivortex pair. The colourmap indicates the value of $m_x$, which corresponds to the direction of the magnetisation in the core (polarisation), showing that the vortex and the antivortex within the loop have the same polarisation.  
b) Mapping the vorticity (represented both by the arrows and the colourmap), reveals that the  loop exhibits a circulating vorticity and is a vortex ring. The vorticity map equally indicates that, in the nearby extended vortex, the vorticity abruptly reverses, corresponding to the presence of a Bloch point. 
Note that the plotted structures have a relatively low vorticity, with $\lvert \Omega \rvert = ~0.1$ (with the exception of the Bloch point). 
c) Plotting preimages for different directions reveals a number of closed loops, that, when the vorticity is plotted, are seen to correspond to vortex rings  (insets). d) In the vicinity of the vortex loop in a), preimages for neighbouring directions are not linked, indicating a Hopf index of zero.}
\end{figure}

Among the plotted structures in Figure~\ref{fig:fig3}, there appear a large number of three-dimensional  `loops', { that resemble }the vortex-antivortex pair schematically illustrated in Figure~\ref{fig:fig1}e. 
We first consider the case of one such loop that is identified using the $\bm{m} = +\mathbf{\hat{x}}$ pre-image\cite{Ackerman2017} in Figure~\ref{fig:fig3}a, where $\bm{m}=\lvert \bm{M}\rvert / {M_s}$ is the reduced magnetisation, and ${M_s}$ is the saturation magnetisation. This loop is located in the vicinity of a single vortex extending throughout the pillar and whose polarisation equally points along the $+\mathbf{\hat{x}}$ direction in the shown slice. 
Considering the magnetisation in the $y-z$ plane, represented by streamlines in Figure \ref{fig:fig3}a, we identify a bound state consisting of two vortices separated by an antivortex, analogous to a cross-tie wall. The loop itself is embedded within a quasi-uniformly magnetised region ($\mathbf{m}=+\mathbf{\hat{x}}$, red) and therefore the vortex and antivortex have same polarisations, as shown schematically in Figure \ref{fig:fig1}e. When the magnetic vorticity vector ${\bm{\Omega}}$ is plotted, see Figure \ref{fig:fig3}b, it  exhibits a unidirectional circulation around the loop, directly comparable to the schematic in Figure \ref{fig:fig1}e. 
This structure is similar to a vortex ring in a fluid, which also corresponds to a loop in the hydrodynamic vorticity. Such vorticity loops have been predicted to exist as  propagating solitons in exchange ferromagnets\cite{Cooper99}. In contrast,  the vortex loops observed here are static and stable at room temperature over the duration of our measurements. We note that the diameter of the vortex ring, i.e. the average distance between the vortex and antivortex cores in the $y-z$ plane, is approximately $370\,\rm{nm}$, and is comparable to the diameter of most other vortex rings present  inside  the pillar. Interestingly, this loop (along with a number of similar vortex rings in the sample) occurs in the vicinity of a singularity: indeed, the neighbouring vortex in the cross-tie structure contains a Bloch point, which can  be located in Figure~\ref{fig:fig3}b where the vorticity, (and the magnetisation in the vortex core) abruptly reverses direction. There is \textit{a priori} no topological requirement for the presence of a Bloch point in proximity of the vortex loop and despite the observed correlations, our static observations do not allow for the determination of a causal relationship between the presence of both structures.

We gain further insight into the topology of these vortex loops by  plotting preimages corresponding to a number of directions of the magnetisation in the vicinity of the vortex-antivortex pair.
The preimage corresponding to the $+\mathbf{\hat{x}}$ direction, i.e. $m_x=+1$, is plotted in light green in Figure \ref{fig:fig3}d, along with additional preimages corresponding to directions indicated in the inset that form an ensemble of closed-loop preimages. The plotted loops do not link, indicating that the vortex ring has a Hopf number $H=0$. Indeed, the vicinity of the $H=0$ structure contains only preimages representing directions close to the $+\mathbf{\hat{x}}$ direction and, consequently, do not cover the $S^2$ sphere, meaning that the magnetisation can  be smoothly unwind into a single point on the sphere\cite{AS2016}. Hence, these vortex rings belong to a class of non-topological solitons \cite{Lee1992}. In the Methods (figure \ref{fig:twoplusone}c), we have developed an analytic model of such a soliton, qualitatively reproducing the observed features, vorticity and pre-images.

\begin{figure}
\centering
\includegraphics[width=0.6\columnwidth]{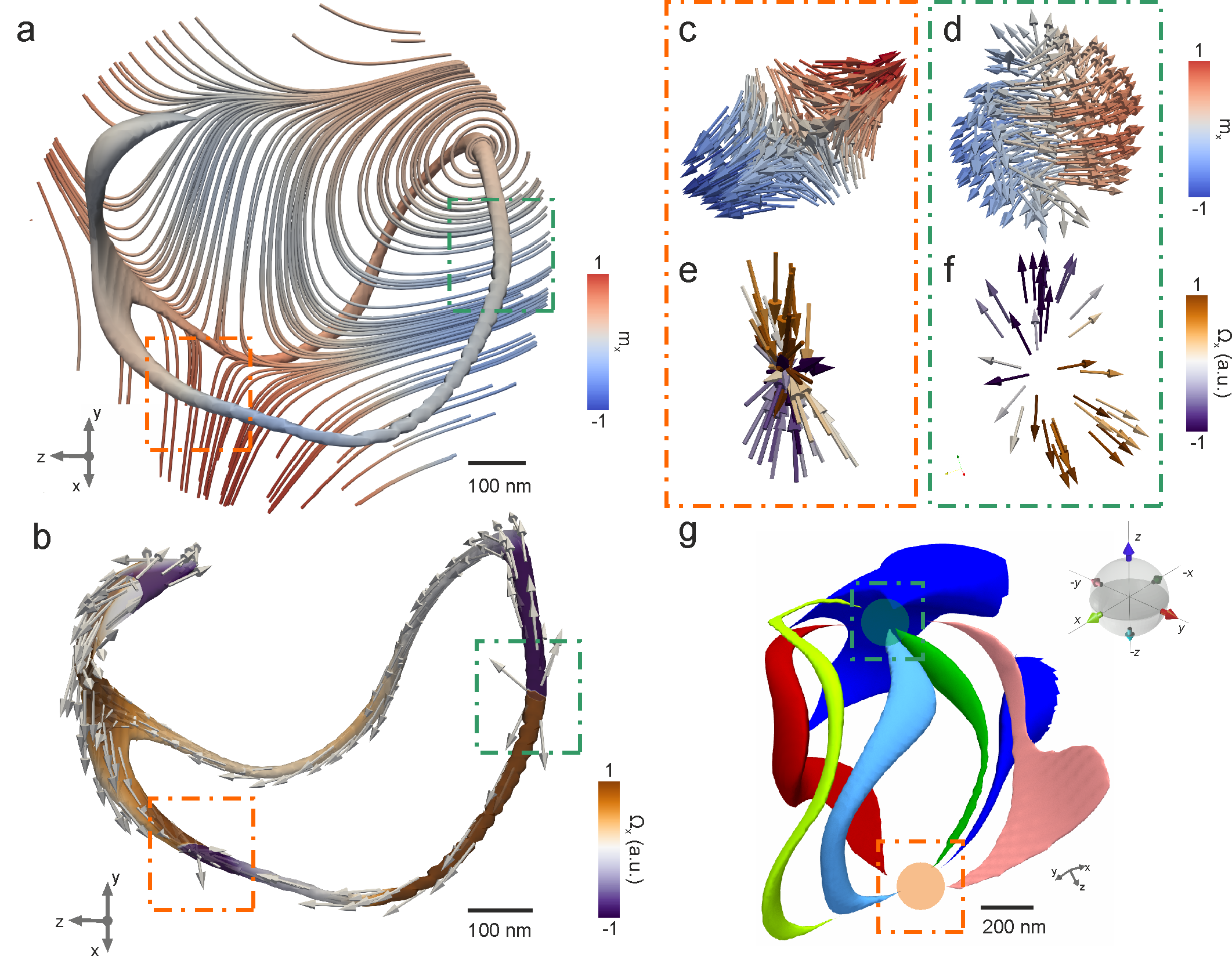}
\caption{\label{fig:fig2} Structure of a vorticity loop containing magnetization singularities. 
a) The vorticity loop is identified by its relatively high magnetic vorticity. 
The magnetic configuration in a two-dimensional slice through the loop is plotted using streamlines to represent the in-plane magnetisation, with the colour indicating the out-of-plane magnetisation component $\pm m_x$ and revealing that the loop consists of a vortex and antivortex pair. 
Within the loop, the $x$ direction of the magnetisation, i.e. the core polarisation, switches from positive (red) to negative (blue) at two points, indicated by the orange and green boxes. b) Plotting the magnetic vorticity reveals that this is in fact not a closed loop, but an ``onion'' state, with the vorticity direction reversing at the same two points. These locations correspond to singularities of the magnetisation (c,d) and, consequently, of the magnetic vorticity (e,f).
g), the preimages corresponding to the Cartesian axes $\pm\mathbf{\hat{x}}$ (light/dark green), $\pm\mathbf{\hat{y}}$ (light/dark red), and $\pm\mathbf{\hat{z}}$ (light/dark blue) are plotted, which reveal an onion-like state, with all preimages meeting at the singularities.}
\end{figure}

In addition to vortex rings, we also identify vorticity loops containing sources and sinks of the magnetisation, due to the presence of Bloch points. The magnetic structure of one such loop is shown in Figure~\ref{fig:fig2}a, where the colourscale indicates the polarisation ($\pm \mathbf{\hat{x}}$) and the magnetisation in a plane of the loop is represented by streamlines, revealing a vortex-antivortex pair. At two points within the loop, the polarisation along the vortex and antivortex cores reverses with the colour changing from blue to red. Consequently, the vorticity does not circulate around the loop, but instead assumes an asymmetric onion-like structure, with the vorticity flowing out from a source (green box in Figure~\ref{fig:fig2}b) and
into a sink (orange box in Figure~\ref{fig:fig2}b). 
The structure of the magnetisation in the vicinity of the singularities is plotted in Figures~\ref{fig:fig2}c,d. In the vicinity of the vorticity sink (Figure \ref{fig:fig2}e), the magnetisation structure (shown in Figure~\ref{fig:fig2}c) corresponds to that of a contra-circulating Bloch point\cite{SM1979} (or anti-Bloch point) with skyrmion number $-1$.
Around the vorticity source (Figure \ref{fig:fig2}f), the magnetisation structure (Figure \ref{fig:fig2}d) corresponds to that of a circulating Bloch\cite{SM1979} point with skyrmion number $+1$. 
Two features of this loop are particularly noteworthy. First, the singularities are not linked to the {generation} and annihilation of a vortex and antivortex with opposite polarisations, as has been reported for dynamic processes\cite{HS2006}. Instead, the pair consists of two half-vortex rings connected by the Bloch points, which locally leads to a reversal of the vorticity along the vortex and the antivortex cores, as seen in Extended Data, Figure \ref{fig:figE2}. Second, while singularities often mediate dynamic processes and have been predicted during magnetisation dynamics\cite{SM1979,Miltat2002} as well as during magnetic field reconnection in plasma physics\cite{Kerr99}, the observed structures are inherently static. In Ref.~\onlinecite{donnelly17}, Bloch points were observed at the locations where a vortex core intersected a domain wall. Similarly, we find that the Bloch point pair is located at the intersection of the vortex-antivortex loop with a domain wall separating regions of opposite  $m_x$.

We gain further insight into the topology of  the vortex-antivortex loop containing singularities by plotting preimages  corresponding to a defined set of directions, or points, on the $S^2$ sphere. In particular, we plot regions of the magnetisation aligned along $\pm \mathbf{\hat{x}}$ (bright/ dark green), $\pm \mathbf{\hat{y}}$ (bright/ dark red), and $\pm \mathbf{\hat{z}}$ (bright/ dark blue) in Figure \ref{fig:fig2}g, which can be seen to form a three-dimensional onion state, with all directions of the magnetisation meeting at the singularities schematically indicated by green and orange circles, corresponding to the anti-Bloch point and Bloch point, respectively. The preimages resemble those found to correspond to `torons', which have recently been observed in chiral liquid crystals \cite{SLCT2009} and anisotropic fluids \cite{LLZ2018}. In the methods, we present an analytical model of different micromagnetic configurations with similar pre-images, allowing us to reproduce and, consequently, understand the experimental observations.

\begin{figure}
\centering
\includegraphics[width=0.7\columnwidth]{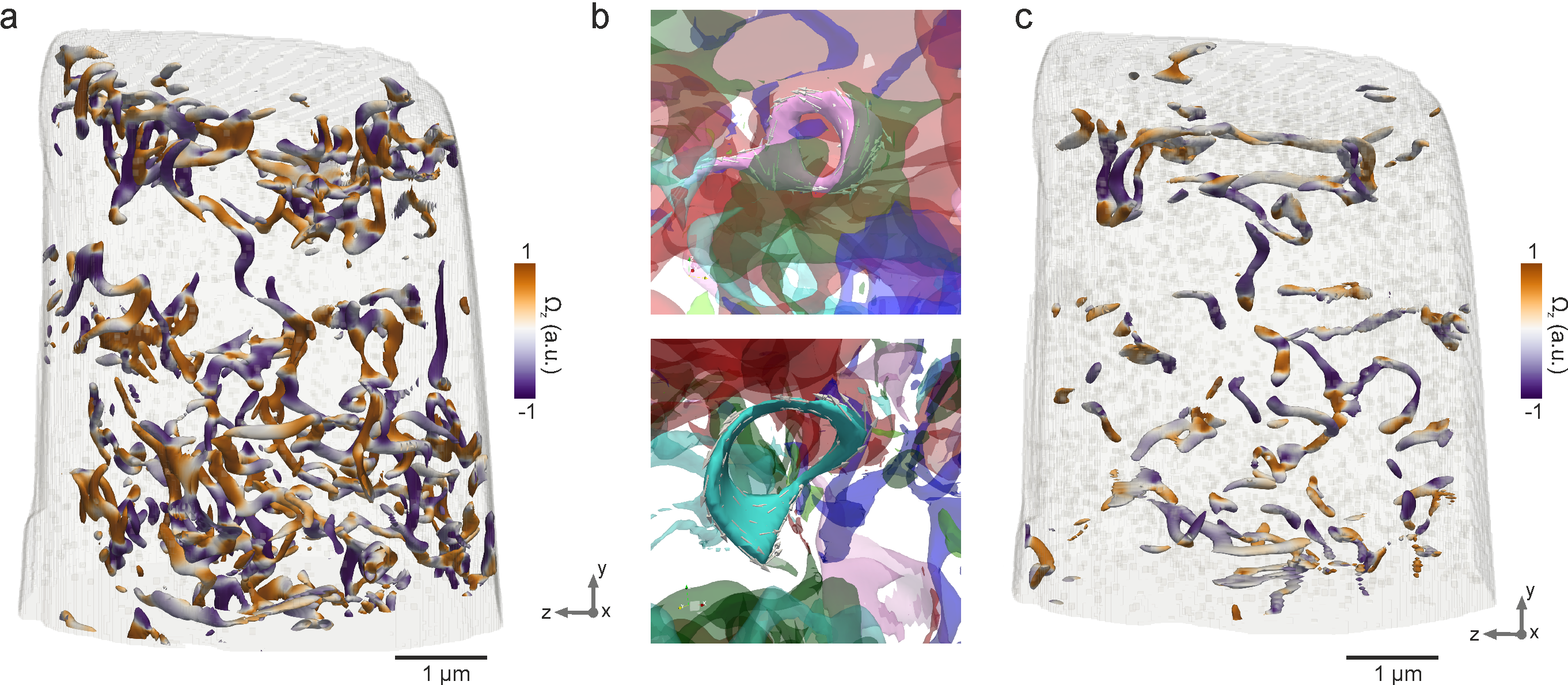}
\caption{\label{fig:fig4} Magnetic vorticity plots measured for the $\rm{GdCo_{2}}$ micropillar at remanence showing the effect of different field histories on the vortex-antivortex structures a) following the application of a 7~T saturating field and c) following saturation and field cooling. A small number of vortex loops like those in figure~\ref{fig:fig3} are present at remanence after the application of a saturating magnetic field, shown in b), however none are observed following the thermal annealing procedure.}
\end{figure}

We explore the stability of the observed vorticity loops by applying two different field and thermal protocols on a similar $\rm{GdCo_2}$ micropillar, and performing magnetic X-ray nanotomography at remanence following each protocol. 
In the first protocol, we apply a $7\,\rm{T}$ magnetic field along the long axis of the pillar at room temperature, and image the resulting remanent configuration. The applied field is above the measured sample saturation field of ca.~$\sim2\,\rm{T}$.
A plot of the magnetic vorticity (see figure~\ref{fig:fig4}a) reveals a large number of vortices and antivortices, as well as magnetic singularities (shown in Methods and figure \ref{fig:ED4} at remanence). By plotting pre-images corresponding to different directions of the magnetisation, we observe a small number of vortex loops, two of which are shown in figure~\ref{fig:fig4}b. The presence of these vortex loops after the application of a saturating magnetic field indicates that the loops can nucleate spontaneously, and therefore do not require a specific field protocol to prepare them.  
Secondly, we heat the sample to $400\,\rm{K}$ while applying a $7\,\rm{T}$ magnetic field. The sample is then field cooled and the field gradually removed after the sample reached room temperature. This annealing procedure is reminiscent of those used to expel defects in single-crystals in order to increase their purity. A plot of the vorticity, shown in figure~\ref{fig:fig4}c, reveals a noticeably smaller number of structures with non-zero vorticity.
Importantly, we do not find any vortex loops, indicating that these are metastable states which are more efficiently destroyed through thermal annealing.
Quantitatively, the average vorticity value  following field cooling is half the value following only the application of a 7~T field, and the total number of Bloch points is roughly halved ($52$ vs. $110$ Bloch points, as seen in figure~\ref{fig:ED4}). 

Although the vortex rings we observe are topologically trivial structures and have a Hopf index of zero, they are surprisingly stable.
We attribute their stability to interactions with  surrounding magnetization structures, which ensure that they are, for example, embedded in larger cross-tie structures or pinned at the intersection with domain walls, {resulting in loops intersected by Bloch points.} 
Moreover, the magnetostatic interaction clearly plays an important role in the stabilisation of these structures, ensuring that our observations of stable localised solitons do not contradict the Hobart-Derrick theorem for an exchange ferromagnet that requires non-linearities (such as intrinsic chirality in the presence of Dzyaloshinskii-Moriya interaction) to set a scale for localised magnetisation non-uniformities. We note that chirality has been demonstrated in a similar bulk amorphous system through the inclusion of structural inhomogeneities\cite{DHKim2019}. We expect that such systems could host topologically non-trivial solitons, such as knots with a higher Hopf number, as well as torons, following predictions for chiral magnetic heterostructures\cite{LLZ2018,Sutcliffe2018,TS2018}, analogous to the reported observations in chiral liquid crystals and ferrofluids\cite{Chen2013,AS2016}. 

Finally, very recent advances in time-resolved X-ray magnetic laminography\cite{donnelly20} open the path to investigating the dynamics of three-dimensional magnetic configurations. As well as probing resonant dynamics, it is possible that investigations of the stability and motion of three-dimensional vortex rings could reveal behaviour analogous to the Kelvin motion of two-dimensional vortex-antivortex pairs \cite{PU1985,Papanicolaou_1999,Cooper98}.  Likewise, we expect that the magnetic vortex loops discovered here containing singularities will also display compelling dynamics, with implications for the fundamental understanding of the role of singularities in magnetic processes.
The study of the conditions for the formation of three-dimensional magnetic structures, and of their stability and  dynamics, is expected to lead to new possibilities for the controlled manipulation of the magnetisation that could be relevant for technological applications requiring complexity, such as neuromorphic computing\cite{huang17} or new proposals for three-dimensional data storage\cite{FSFHFC2017}. 

\section{Methods}
\renewcommand\thefigure{M\arabic{figure}}
\setcounter{figure}{0}

\subsection{Sample Fabrication}

The samples investigated were both $\rm{GdCo_2}$ micropillars of diameter $5\,\rm{\mu m}$ that were cut from a larger nugget of $\rm{GdCo_2}$ using a focused ion beam in combination with a micromanipulator, and mounted on top of OMNY tomography pins\cite{hollerpin17}.
\subsection{X-ray ptychographic tomography}
Hard X-ray magnetic tomography was performed at the cSAXS beamline at the Swiss Light Source, Paul Scherrer Institut, using the flexible tomographic nano imaging (flOMNI) instrument\cite{holler17}. Part of the data presented in this manuscript (the central vortex containing the Bloch point in Figure \ref{fig:fig3}a,b) formed part of the dataset presented in Ref. \onlinecite{donnelly17}. All other data is shown and analysed for the first time here.

Two dimensional tomographic projections were measured with X-ray ptychography, a coherent diffractive imaging technique allowing access to the full complex transmission function of the sample\cite{ptychoreview,rodenburg07}.  
For X-ray ptychography, an X-ray illumination of approximately $4\,\rm{\mu m}$ was defined on the sample, and ptychography scans were performed by measuring diffraction patterns on a concentric grid of circles with a radial separation of $0.4\,\rm{\mu m}$ for a field of view of $8\times7\,\rm{\mu m}$ and $13\times9\,\rm{\mu m}$ for the untilted and tilted sample orientation, respectively. The projections were reconstructed using 500 iterations of the difference map and 200 iterations of the maximum likelihood refinement using the cSAXS PtychoShelves package \cite{Wakonig20}.

To probe the magnetisation of the sample, X-rays tuned to the Gd $L_3$ edge with a photon energy of $7.246\,\rm{keV}$ were chosen to maximise the absorption XMCD signal\cite{Donnelly2016}. Circularly polarised X-rays were produced by including a $500\,\rm{\mu m}$-thick diamond phase plate upstream of the sample position\cite{Scagnoli09}. The degree of circular polarisation achieved was greater than $99\%$, and with an transmission of approximately $35\%$. 

The tomographic projections were aligned with high precision as described in Ref.~\cite{donnelly17}.

\subsection{Magnetic tomography}
When a single circular polarisation projection is measured, the component of the magnetisation parallel to the X-ray beam is probed, along with the electronic structure of the sample. To probe all three components of the magnetisation, projections were measured around a rotation axis for two orientations of the sample\cite{donnelly17}.  Generally, the magnetic contrast of a projection is isolated from other contrast mechanisms by measuring the same projection using circular left and right polarised light, where the sign of the magnetic contrast is reversed, and taking the difference between the two images. Here, a single X-ray  polarisation is used for all measurements and, in order to isolate the magnetic structure, projections with circularly left polarisation are measured at $\theta$ and $\theta+ 180^\circ$. Between these two angles, the magnetic contrast is reversed, which can be used to differentiate the magnetic contrast from the electronic contrast. Therefore, for the magnetic tomography measurements, circular left polarisation projections were measured through $360^\circ$ about the rotation axis, instead of through $180^\circ$, as in standard tomography. 

The magnetisation {(which is a three-dimensional vector field)} was reconstructed using a two-step gradient-based iterative reconstruction algorithm, described in Ref.~\cite{Donnelly_thesis}. The spatial resolution for each component of the magnetisation was estimated using Fourier Shell Correlation\cite{vanheel05}, and a three-dimensional Hanning low-pass filter was used  to remove high-frequency noise.  The spatial resolution of the reconstructed magnetisation was found to be $97\,\rm{nm}$, $125\,\rm{nm}$ and $127\,\rm{nm}$ in the $x-z$, $x-y$ and $y-z$ planes, respectively\cite{donnelly17}.

The magnetic vorticity was calculated according to Equation \ref{eq:magvort}. The magnetisation was normalised to obtain the unit length, which was used to calculate the magnetic vorticity. The three-dimensional visualisations of the magnetic vorticity and magnetisation were performed with Paraview. 

To consider the topology of the magnetisation in three dimensions, pre-images corresponding to different directions are plotted within the pillar. The difference between the magnetisation vector and the $m_x=1$ direction is calculated using:
\begin{equation}
\delta_{px} = \left(\frac{m_x}{\lvert \boldsymbol{m}\rvert}-1\right)^2+\left(\frac{m_y}{\lvert \boldsymbol{m}\rvert}\right)^2+\left(\frac{m_z}{\lvert \boldsymbol{m}\rvert}\right)^2
\end{equation}
To plot the $m_x=1$ pre-image, we plot an isosurface for $\delta_{px}=0.01$. This results in a tube rather than a line, which is necessary due to the finite spatial resolution and signal-to-noise ratio of the measurement. 

\subsection{Analytical models}

{To qualitatively interpret and understand the observed structures, we build a series of 2+1 dimensional models, which allow comparing the observed magnetization structures, preimages and the vorticity with the ones derived from modeled vortex loops with different magnetization structures.} 
{These models are} similar to those used for description of hopfions in Ref.~\onlinecite{WZ1983}. They are based on the subdivision of the magnetic material volume into thin slices, lying in the $x-y$ plane of a Cartesian coordinate system. The magnetisation in each slice can then be described by a complex function $w$ of a complex variable $u=x+\imath y$ by means of stereographic projection $\{\umx+\imath \umy, \umz\} = \{ 2 w, 1-w\wc \}/(1+w\wc)$, where the over-line denotes complex conjugation, so that $\uc=x - \imath y$, $\imath=\sqrt{-1}$. Without loss of generality, any three-dimensional magnetisation distribution $\vec{m}(x, y, z)$ can be described by a function $w=w(u,\uc,z)$, which depends on the complex coordinate $u$ within each slice and the extra-dimensional variable $z$, identifying the slice.

For realistic models, including at least the exchange and the magnetostatic interactions, no exact solutions for non-uniform $w(u,\uc,z)$ are known. However, if the magnetostatic interaction is neglected and $w(u,\uc,z)$ is assumed to be weakly dependent on $z$, two large families of exact solutions exist for $w(u,\uc,z)$ at a fixed $z$. These are solitons\cite{BP75}, which are meromorphic functions $w(u,\uc,z)=f(u,z)$, and singular merons\cite{G78}, which are functions with $|w(u,\uc,z)|=1$ or $w(u,\uc,z)=f(u,z)/|f(u,z)|$. Zeros of $f(u,z)$ correspond to the centers of magnetic vortices (or hedgehog-like structures, if the magnetisation vectors are rotated by $\pi/2$ in the $x$-$y$ plane). The poles correspond to the centers of the magnetic antivortices (or saddles). From the stereographic projection it follows that for solitons $m_z=1$ in the centers of the vortices and $m_z=-1$ in the centers of antivortices.

An example of meromorphic functions are the rational functions of a complex argument (quotient of two polynomials). They allow direct expression of the vortex/antivortex pair annihilation as a cancellation of two identical monomials, whereas creation is a time-reversed process. The number of vortices in each slice is a conserved quantity\cite{BP75} (topological charge, or skyrmion number) in the sense that it cannot be changed by a smooth singularity-free variation of the magnetisation distribution. For the slices in the $x$-$y$ plane the topological charge density is the $z$-component of the vorticity $\Omega_z$ and the total charge is the integral of this density over the whole slice. Creation and annihilation of the vortex-antivortex pairs within the soliton is always accompanied by a singularity.

\begin{figure}
\centering
\includegraphics[width=0.9\columnwidth]{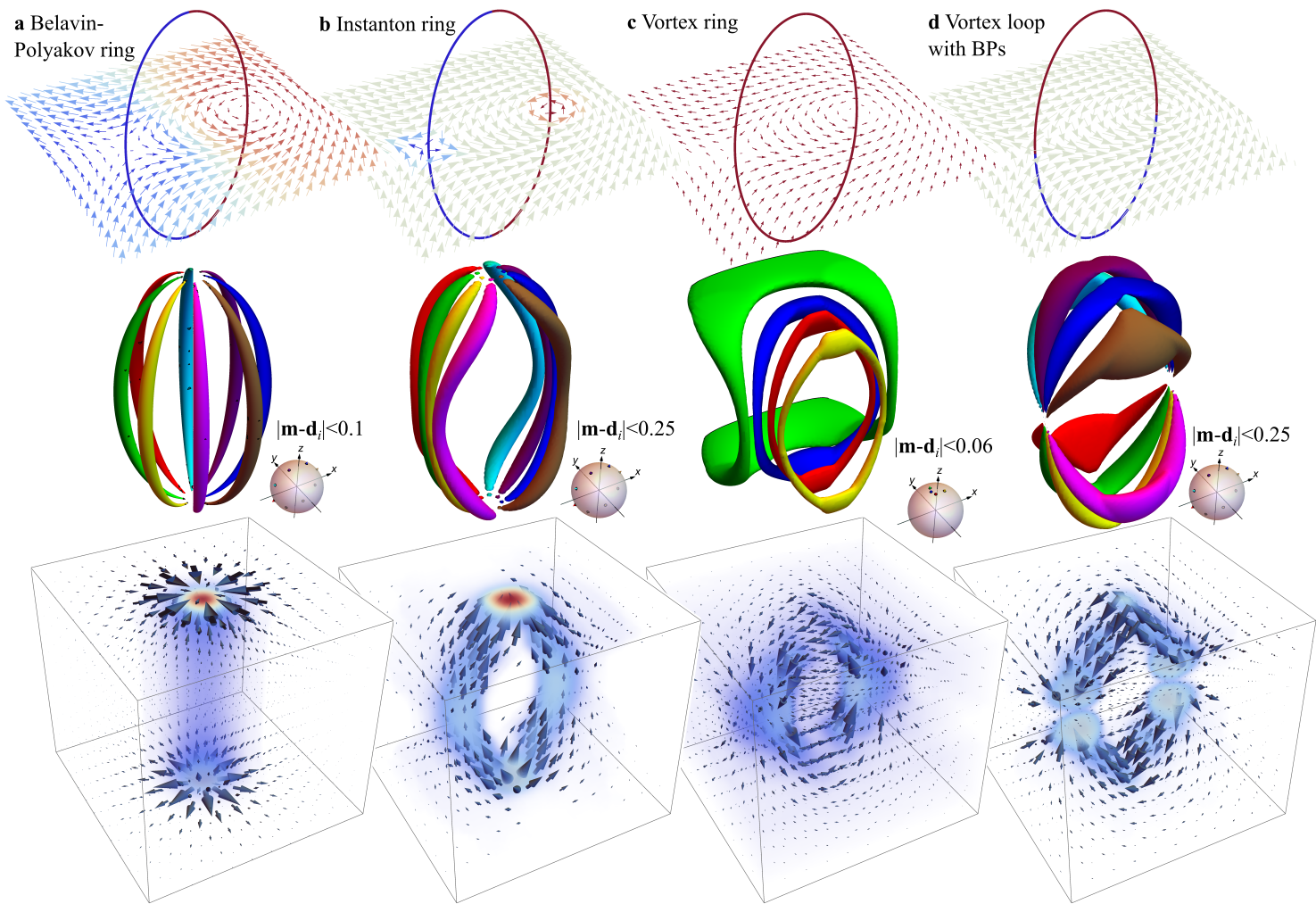}
\caption{\label{fig:twoplusone}Top to bottom:  Magnetisation, pre-images and vorticity distribution for the different $2+1$ dimensional analytical models, discussed in the methods section. The magnetisation plots only {include} the projection of the magnetisation onto the shown planes, {while} the rings correspond to the positions of the vortex and antivortex centers, and the color indicates the $m_\mathrm{Z}$ component of the magnetisation. The preimages are shown as volumes where the magnetisation vectors deviate only slightly from certain directions $\vec{d}_i$, {indicated} by the color-coded arrows on each corresponding legend. The opacity and color on the vorticity plots indicates the magnitude of local vorticity vectors. The structure in c {is comparable to the} vortex rings in figure~\ref{fig:fig3}, while the structure in d {is comparable to that} in figure~\ref{fig:fig2}.}
\end{figure}

\begin{figure}
\centering
\includegraphics[width=\columnwidth]{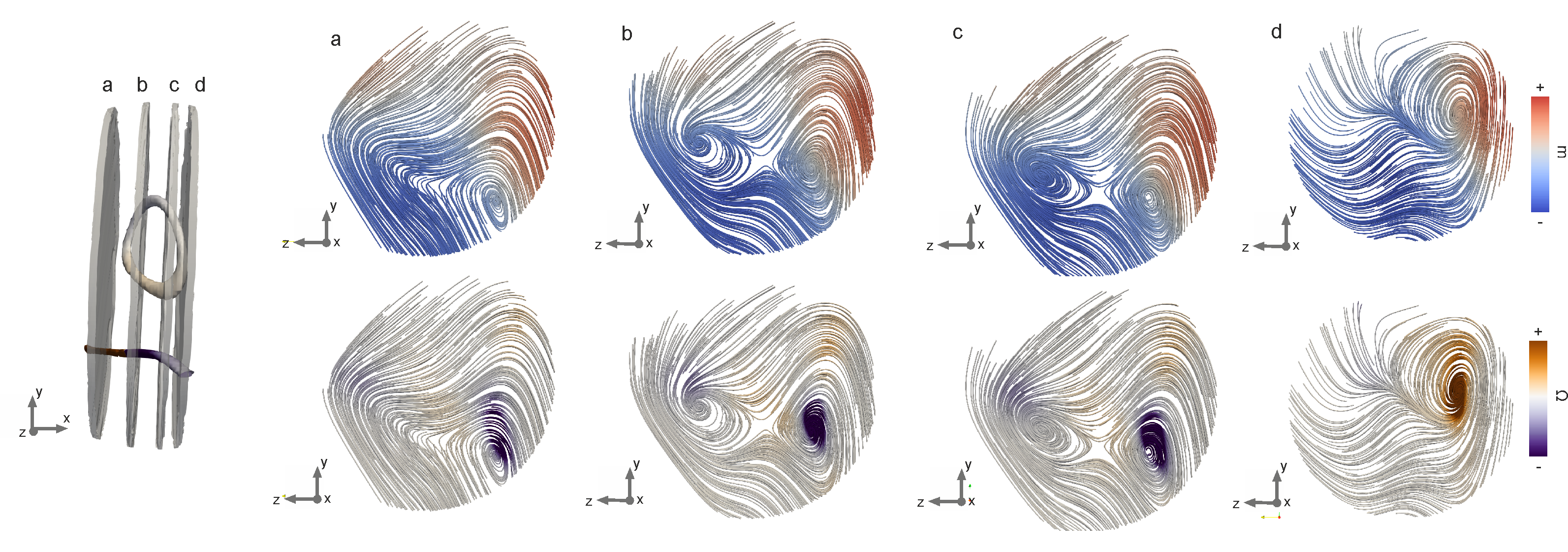}
\caption{\label{fig:figE2} Detailed overview of the vortex ring, shown in successive slices through the loop. The colourscale in the top row indicates the magnetisation, while the colourscale in the bottom row indicates the vorticity. The vorticity associated with the vortex structure extending throughout the pillar changes in sign in slice \emph{d} due to the presence of a Bloch point, while the vortex-antivortex pair conserves its vorticity throughout. In slices \emph{b} and \emph{c}, the magnetisation forms a cross-tie wall like structure, which dissolves as the pair unwinds, at slices \emph{a} and \emph{d}, leaving the a single vortex. }
\end{figure}

A vortex ring can be understood as a process of creation, separation, convergence and annihilation of a vortex-antivortex pair as the variable $z$ advances through the successive slices\cite{Cooper99}. Consider
\begin{equation}
 \label{eq:BPring}
 w_\mathrm{BPr}(u,\uc,z) = f(u,z)=\imath \frac{u - p(z)}{u + p(z)}=\imath \frac{u - \sqrt{1-(z/2)^2}}{u + \sqrt{1-(z/2)^2}}
\end{equation}
for an (arbitrary) range $-2<z<2$, where the specific expression for $p(z)$ was chosen to make the vortex and antivortex cores extend along arcs, as in the experimental data. It describes the creation of a vortex-antivortex pair at $x=y=0$ and $z=2$, the vortex and antivortex moving apart (with the maximum distance between their centres equal to $2$ at $z=0$), then approaching each other again, and annihilating at $z=-2$. We call this model the Belavin-Polyakov ring because each slice is a Belavin-Polyakov soliton, described by a meromorphic $w(u,\uc,z)$. The corresponding schematic magnetisation, set of preimages and vorticity are shown in figure \ref{fig:twoplusone}a. A similar preimage patterns connecting two Bloch points were indeed observed in our sample. However, the corresponding vorticity distributions are different. Indeed, instead of a single centrally-symmetric vorticity bundle we reconstruct a pair of bundles, corresponding to the vortex and antivortex centers. Clearly, the pure Belavin-Polyakov ring model can not reproduce this feature.

To 'unbundle' the vortex and antivortex, we can use the instanton model\cite{G78} by writing:
\begin{equation}
 \label{eq:instanton-ring}
 w_\mathrm{i}(u,\uc,z) =
 \begin{cases} 
      f(u,z)/c(z) & |f(u,z)| \leq c(z) \\
      f(u,z)/|f(u,z)| & d(z) > |f(u,z)| > c(z) \\
      f(u,z)/d(z) & |f(u,z)| >d(z)
 \end{cases},
\end{equation}
where $d(z)=1/c(z)$, assuming the same size for the vortex and antivortex cores. Choosing $c(z)=1-q+q|z|/2<1$ allows the control of the size of the vortex and antivortex cores (where $m_z\neq0$) at the central plane $z=0$ via the parameter $q$. The magnetisation, preimages and vorticity for such an instanton ring with $q=3/4$ are shown in figure~\ref{fig:twoplusone}b. While they reproduce qualitatively both the vorticity distribution and the preimages, shown in figures~\ref{fig:fig2}b and \ref{fig:fig2}g, the structure of the Bloch points is different. Indeed, the instanton ring has two hedgehog-type Bloch points (in which the  magnetisation directions are opposite), whereas the observed structure, shown in figure~\ref{fig:fig2}, contains two different types of Bloch points. {Additionally, this model differs from the observation in figure~\ref{fig:fig2} in that singularities are absent at the transition from the experimentally-observed vortex and antivortex pair to a uniformly-magnetized region.}
The Bloch points in figure~\ref{fig:fig2} {rather coincide} with the polarisation reversal of vortex and antivortex cores as they propagate through the volume of the sample. {In order to analytically describe this structure, we first need to build a model for a vortex ring.}

{To describe a vortex-antivortex pair unbound by Bloch point singularities, the vortex and the antivortex must have identical polarisations (i.e. the same direction of $m_z$ within the core).}
In this case the topological charge in each slice is zero. 
Such a configuration can be obtained as a generalisation of (\ref{eq:instanton-ring})
\begin{equation}
 \label{eq:quasiuniform-ring}
 w_\mathrm{r}(u,\uc,z) = A(z)
 \begin{cases} 
      f(u,z)/c(z) & |f(u,z)| \leq c(z) \\
      f(u,z)/|f(u,z)| & d(z) > |f(u,z)| > c(z) \\
      d(z)/\overline{f(u,z)} & |f(u,z)| >d(z)
 \end{cases},
\end{equation}
where the modification to the last line reverses the polarisation of the antivortex. The factor $A(z)=(1-z^2/4)s$ ensures that, at $z=\pm 2$, the function $w_\mathrm{r} = 0$, which corresponds to the uniform state. The parameter $s$ allows for the control of the degree of quasiuniformity: the smaller $s$ is, the less $m_z$ deviates from $1$. The magnetisation, preimages and vorticity for such a quasiuniform ring with $q=3/4$ and $s=1/4$ are shown in figure~\ref{fig:twoplusone}c. They are qualitatively analogous to the experimentally-observed vortex rings in figures~\ref{fig:fig3}b and \ref{fig:fig3}d.

Finally, we can {extend the above model to} a vortex ring in which the polarisation reverses along the vortex and the antivortex cores, {in the presence of Bloch points}. To describe this state, {we} note that with $s=1$, $c(z)=z^2/4$,  the magnetisation of the quasiuniform ring (\ref{eq:quasiuniform-ring}) at $z=0$ lies completely in the $x$-$y$ plane except for at the centres of the the vortex and antivortex, where its direction is undefined. Joining at the central plane two half-rings with opposite polarisations:
\begin{equation}
 \label{eq:reversal-ring}
 w_\mathrm{vls}(u,\uc,z) = A(z)
 \begin{cases} 
       w_\mathrm{r}(u,\uc,z) & z \leq 0 \\
      1/\overline{w_\mathrm{r}(u,\uc,z)} & z>0
 \end{cases}
\end{equation}
yields the model for the vortex loop with Bloch point singularities, shown in figure~\ref{fig:twoplusone}d. The structure corresponds well to the observations in figure~\ref{fig:fig2}, including the observed Bloch point types.

Note that despite piecewise nature of the above functions, the resulting magnetisation vector fields are continuous (apart at the Bloch points). While neither ansatz in the presented series is an exact solution of the corresponding micromagnetic problem (not even of its restricted exchange-only version), they provide a simple and easily interpretable model to understand the observed magnetisation distributions.

\begin{figure}
\centering
\includegraphics[width=\columnwidth]{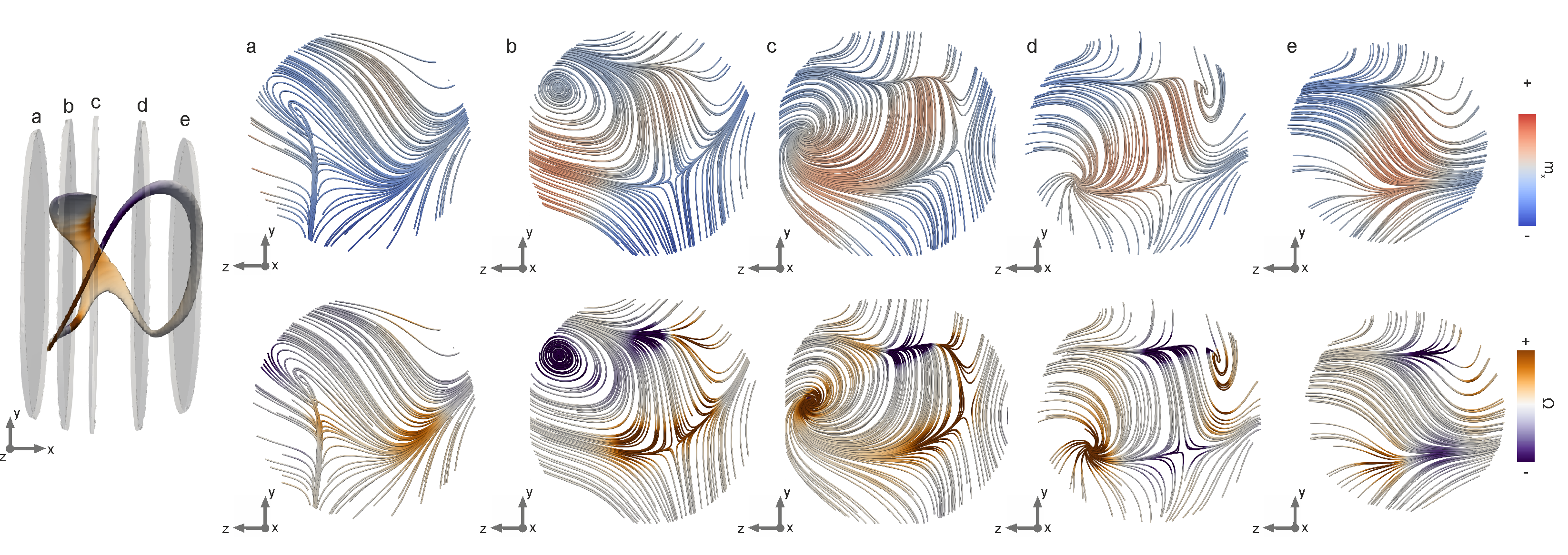}
\caption{\label{fig:figE3} Detailed overview of the magnetic state of the vortex loop containing Bloch points, shown in successive slices through the loop. The colourscale in the top row indicates the magnetisation, while the colourscale in the bottom row indicates the vorticity. The vorticity along the vortex core reverses between slices \emph{b} and \emph{c}, while the vorticity along the antivortex core reverses between slices \emph{c} and \emph{d}. 
}
\end{figure}

\begin{figure}
\centering
\includegraphics[width=0.7\columnwidth]{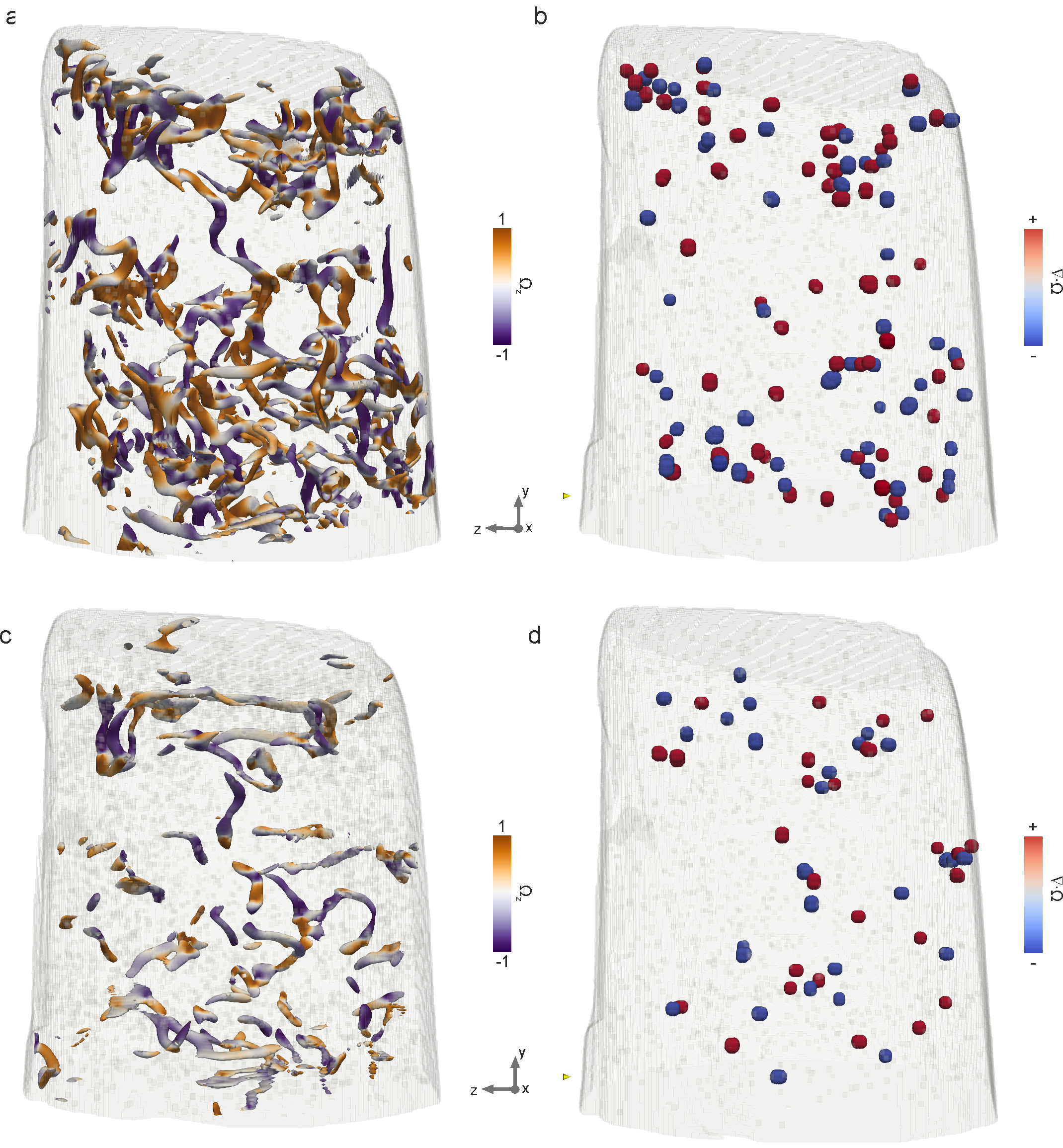}
\caption{\label{fig:ED4} Plotting regions of significant magnetic vorticity reveals the effect of different field histories on the network of vortex-antivortex structures present in a GdCo$_\mathrm{2}$ micropillar  
a) following the application of a 7~T saturating field and c) following saturation and field cooling. Regions of high divergence of the magnetic vorticity indicate the presence of Bloch points (red) and anti-Bloch points (blue) (b) at remanence, following saturation and d) after heating at 400 K and field cooling in a 7~T field. Noticeably fewer magnetic structures with high vorticity are present after the field cooling procedure in than after the simple application of a magnetic field.
}
\end{figure}

\section{Contributions}
The study of topological magnetic features in three dimensions was conceived by S.G., C.D. and K.L.M., and originated from a larger project on three-dimensional magnetic systems conceived by L.J.H and J.R..  C.D., M.G.-S., S.G., V.S., M.H. and J.R. performed the experiments. Magnetometry measurements of the material were performed by N.S.B..  C.D. performed the magnetic reconstruction with support from M.G.-S. and V.S.. C.D. analysed the data and N.C. conceived the calculation of the magnetic vorticity. C.D., S.G., K.L.M. and N.C. interpreted the magnetic configuration.  K.L.M. developed the analytical model. C.D., K.L.M., N.C. and S.G. wrote the manuscript with  contributions from all authors. 
\section{Acknowledgements}
X-ray measurements were performed at the cSAXS beamline at the Swiss Light Source, Paul Scherrer Institut, Switzerland. The authors are grateful to Andrei Bogatyr\"ev for his careful reading of the manuscript and many valuable remarks. We thank R. M. Galera for providing and performing magnetic characterisations of the GdCo$_2$ nugget,  S. Stutz for the sample fabrication, and E. M\"uller from the Electron Microscopy Facility at PSI for the FIB-preparation of the pillar samples. C.D.  is supported by the Leverhulme Trust (ECF-2018-016), the Isaac Newton Trust (18-08) and the L’Oréal-UNESCO UK and Ireland Fellowship For Women In Science. S.G. was funded by the Swiss National Science Foundation, Spark Project Number 190736. 
K.L.M. acknowledges the support of the Russian Science Foundation under the project RSF~16-11-10349. N.C. was supported by EPSRC Grant EP/P034616/1 and by a Simons Investigator Award.

\section{Competing interests}
The authors declare no competing financial interests.

\section{Corresponding authors}
Correspondence to C.D., K.L.M. or S.G.

\section{Data and Code Availability}
All data and codes will be made available on a repository following the publication of the manuscript.

\bibliographystyle{naturemag}

\begin{thebibliography}{10}
\expandafter\ifx\csname url\endcsname\relax
  \def\url#1{\texttt{#1}}\fi
\expandafter\ifx\csname urlprefix\endcsname\relax\def\urlprefix{URL }\fi
\providecommand{\bibinfo}[2]{#2}
\providecommand{\eprint}[2][]{\url{#2}}

\bibitem{Yao1996}
\bibinfo{author}{Yao, J.} \& \bibinfo{author}{Lundgren, T.}
\newblock \bibinfo{title}{Experimental investigation of microbursts}.
\newblock \emph{\bibinfo{journal}{Experiments in Fluids}}
  \textbf{\bibinfo{volume}{21}}, \bibinfo{pages}{17 -- 25}
  (\bibinfo{year}{1996}).

\bibitem{Kilner2000}
\bibinfo{author}{Kilner, P.~J.} \emph{et~al.}
\newblock \bibinfo{title}{Asymmetric redirection of flow through the heart}.
\newblock \emph{\bibinfo{journal}{Nature}} \textbf{\bibinfo{volume}{404}},
  \bibinfo{pages}{759 -- 761} (\bibinfo{year}{2000}).
\newblock \bibinfo{note}{Article}.

\bibitem{Stenhoff-book}
\bibinfo{author}{Stenhoff, M.}
\newblock \emph{\bibinfo{title}{{Ball Lightning: An Unsolved Problem in
  Atmospheric Physics}}} (\bibinfo{publisher}{Springer-Verlag},
  \bibinfo{address}{US}, \bibinfo{year}{1999}), \bibinfo{edition}{1} edn.

\bibitem{Akhmetov-book}
\bibinfo{author}{Akhmetov, D.~G.}
\newblock \emph{\bibinfo{title}{{Vortex Rings}}}
  (\bibinfo{publisher}{Springer-Verlag}, \bibinfo{address}{Berlin, Heidelberg},
  \bibinfo{year}{2009}), \bibinfo{edition}{1} edn.

\bibitem{Cooper99}
\bibinfo{author}{Cooper, N.~R.}
\newblock \bibinfo{title}{{Propagating Magnetic Vortex Rings in Ferromagnets}}.
\newblock \emph{\bibinfo{journal}{Phys. Rev. Lett.}}
  \textbf{\bibinfo{volume}{82}}, \bibinfo{pages}{1554--1557}
  (\bibinfo{year}{1999}).

\bibitem{donnelly17}
\bibinfo{author}{Donnelly, C.} \emph{et~al.}
\newblock \bibinfo{title}{{Three-dimensional magnetization structures revealed
  with X-ray vector nanotomography}}.
\newblock \emph{\bibinfo{journal}{Nature}} \textbf{\bibinfo{volume}{547}},
  \bibinfo{pages}{328--331} (\bibinfo{year}{2017}).

\bibitem{Feldtkeller1965}
\bibinfo{author}{Feldtkeller, E.}
\newblock \bibinfo{title}{{Mikromagnetisch stetige und unstetige
  magnetisierungskonfigurationen}}.
\newblock \emph{\bibinfo{journal}{Zeitschrift f{\"u}r angewandte Physik}}
  \textbf{\bibinfo{volume}{19}}, \bibinfo{pages}{530--536}
  (\bibinfo{year}{1965}).

\bibitem{SOHSO00}
\bibinfo{author}{Shinjo, T.}, \bibinfo{author}{Okuno, T.},
  \bibinfo{author}{Hassdorf, R.}, \bibinfo{author}{Shigeto, K.} \&
  \bibinfo{author}{Ono, T.}
\newblock \bibinfo{title}{{Magnetic Vortex Core Observation in Circular Dots of
  Permalloy}}.
\newblock \emph{\bibinfo{journal}{Science}} \textbf{\bibinfo{volume}{289}},
  \bibinfo{pages}{930--932} (\bibinfo{year}{2000}).

\bibitem{WWBPMW02}
\bibinfo{author}{Wachowiak, A.} \emph{et~al.}
\newblock \bibinfo{title}{{Direct observation of internal spin structure of
  magnetic vortex cores.}}
\newblock \emph{\bibinfo{journal}{{Science}}} \textbf{\bibinfo{volume}{298}},
  \bibinfo{pages}{577--580} (\bibinfo{year}{2002}).

\bibitem{G2008}
\bibinfo{author}{Guslienko, K.}
\newblock \bibinfo{title}{{Magnetic vortex state stability reversal and
  dynamics in restricted geometries}}.
\newblock \emph{\bibinfo{journal}{Journal of nanoscience and nanotechnology}}
  \textbf{\bibinfo{volume}{8}}, \bibinfo{pages}{2745--60}
  (\bibinfo{year}{2008}).

\bibitem{Choe2004}
\bibinfo{author}{Choe, S.-B.} \emph{et~al.}
\newblock \bibinfo{title}{Vortex core-driven magnetization dynamics}.
\newblock \emph{\bibinfo{journal}{Science}} \textbf{\bibinfo{volume}{304}},
  \bibinfo{pages}{420--422} (\bibinfo{year}{2004}).
\newblock \urlprefix\url{https://science.sciencemag.org/content/304/5669/420}.
\newblock
  \eprint{https://science.sciencemag.org/content/304/5669/420.full.pdf}.

\bibitem{Waeyenberge2006}
\bibinfo{author}{Van~Waeyenberge, B.} \emph{et~al.}
\newblock \bibinfo{title}{Magnetic vortex core reversal by excitation with
  short bursts of an alternating field}.
\newblock \emph{\bibinfo{journal}{Nature}} \textbf{\bibinfo{volume}{444}},
  \bibinfo{pages}{461 -- 464} (\bibinfo{year}{2006}).
\newblock \bibinfo{note}{Article}.

\bibitem{HGFS07}
\bibinfo{author}{Hertel, R.}, \bibinfo{author}{Gliga, S.},
  \bibinfo{author}{F{\"a}hnle, M.} \& \bibinfo{author}{Schneider, C.~M.}
\newblock \bibinfo{title}{{Ultrafast Nanomagnetic Toggle Switching of Vortex
  Cores}}.
\newblock \emph{\bibinfo{journal}{Phys. Rev. Lett.}}
  \textbf{\bibinfo{volume}{98}}, \bibinfo{pages}{117201}
  (\bibinfo{year}{2007}).

\bibitem{Pigeau2010}
\bibinfo{author}{Pigeau, B.} \emph{et~al.}
\newblock \bibinfo{title}{A frequency-controlled magnetic vortex memory}.
\newblock \emph{\bibinfo{journal}{Applied Physics Letters}}
  \textbf{\bibinfo{volume}{96}}, \bibinfo{pages}{132506}
  (\bibinfo{year}{2010}).
\newblock \urlprefix\url{https://doi.org/10.1063/1.3373833}.
\newblock \eprint{https://doi.org/10.1063/1.3373833}.

\bibitem{HS2006}
\bibinfo{author}{Hertel, R.} \& \bibinfo{author}{Schneider, C.~M.}
\newblock \bibinfo{title}{{Exchange Explosions: Magnetization Dynamics during
  Vortex-Antivortex Annihilation}}.
\newblock \emph{\bibinfo{journal}{Phys. Rev. Lett.}}
  \textbf{\bibinfo{volume}{97}}, \bibinfo{pages}{177202}
  (\bibinfo{year}{2006}).

\bibitem{gliga2008}
\bibinfo{author}{Gliga, S.}, \bibinfo{author}{Yan, M.},
  \bibinfo{author}{Hertel, R.} \& \bibinfo{author}{Schneider, C.~M.}
\newblock \bibinfo{title}{{Ultrafast dynamics of a magnetic antivortex:
  Micromagnetic simulations}}.
\newblock \emph{\bibinfo{journal}{Phys. Rev. B}} \textbf{\bibinfo{volume}{77}},
  \bibinfo{pages}{060404} (\bibinfo{year}{2008}).

\bibitem{gligaJAP2008}
\bibinfo{author}{Gliga, S.}, \bibinfo{author}{Hertel, R.} \&
  \bibinfo{author}{Schneider, C.~M.}
\newblock \bibinfo{title}{Switching a magnetic antivortex core with ultrashort
  field pulses}.
\newblock \emph{\bibinfo{journal}{Journal of Applied Physics}}
  \textbf{\bibinfo{volume}{103}}, \bibinfo{pages}{07B115}
  (\bibinfo{year}{2008}).

\bibitem{Neudert2006}
\bibinfo{author}{Neudert, A.} \emph{et~al.}
\newblock \bibinfo{title}{Bloch-line generation in cross-tie walls by fast
  magnetic-field pulses}.
\newblock \emph{\bibinfo{journal}{Journal of Applied Physics}}
  \textbf{\bibinfo{volume}{99}}, \bibinfo{pages}{08F302}
  (\bibinfo{year}{2006}).
\newblock \urlprefix\url{https://doi.org/10.1063/1.2170399}.
\newblock \eprint{https://doi.org/10.1063/1.2170399}.

\bibitem{BP75}
\bibinfo{author}{Belavin, A.~A.} \& \bibinfo{author}{Polyakov, A.~M.}
\newblock \bibinfo{title}{{Metastable states of two-dimensional isotropic
  ferromagnet.}}
\newblock \emph{\bibinfo{journal}{ZETP lett.}} \textbf{\bibinfo{volume}{22}},
  \bibinfo{pages}{245--247} (\bibinfo{year}{1975}).

\bibitem{Senthil2004}
\bibinfo{author}{Senthil, T.}, \bibinfo{author}{Vishwanath, A.},
  \bibinfo{author}{Balents, L.}, \bibinfo{author}{Sachdev, S.} \&
  \bibinfo{author}{Fisher, M. P.~A.}
\newblock \bibinfo{title}{Deconfined quantum critical points}.
\newblock \emph{\bibinfo{journal}{Science}} \textbf{\bibinfo{volume}{303}},
  \bibinfo{pages}{1490--1494} (\bibinfo{year}{2004}).
\newblock \urlprefix\url{https://science.sciencemag.org/content/303/5663/1490}.
\newblock
  \eprint{https://science.sciencemag.org/content/303/5663/1490.full.pdf}.

\bibitem{Ackerman2017}
\bibinfo{author}{Ackerman, P.~J.} \& \bibinfo{author}{Smalyukh, I.~I.}
\newblock \bibinfo{title}{{Diversity of Knot Solitons in Liquid Crystals
  Manifested by Linking of Preimages in Torons and Hopfions}}.
\newblock \emph{\bibinfo{journal}{Phys. Rev. X}} \textbf{\bibinfo{volume}{7}},
  \bibinfo{pages}{011006} (\bibinfo{year}{2017}).

\bibitem{Donnelly2016}
\bibinfo{author}{Donnelly, C.} \emph{et~al.}
\newblock \bibinfo{title}{{High-resolution hard x-ray magnetic imaging with
  dichroic ptychography}}.
\newblock \emph{\bibinfo{journal}{Phys. Rev. B}} \textbf{\bibinfo{volume}{94}},
  \bibinfo{pages}{064421} (\bibinfo{year}{2016}).

\bibitem{Donnelly2018}
\bibinfo{author}{Donnelly, C.} \emph{et~al.}
\newblock \bibinfo{title}{{Tomographic reconstruction of a three-dimensional
  magnetization vector field}}.
\newblock \emph{\bibinfo{journal}{New Journal of Physics}}
  \textbf{\bibinfo{volume}{20}}, \bibinfo{pages}{083009}
  (\bibinfo{year}{2018}).

\bibitem{chikazumi-book}
\bibinfo{author}{Chikazumi, S.}
\newblock \emph{\bibinfo{title}{Physics of ferromagnetism}},
  vol.~\bibinfo{volume}{94} of \emph{\bibinfo{series}{International Series of
  Monographs on Physics}} (\bibinfo{publisher}{Oxford University Press},
  \bibinfo{address}{Oxford ; New York}, \bibinfo{year}{2010}),
  \bibinfo{edition}{2} edn.

\bibitem{AHA79}
\bibinfo{author}{Arrott, A.}, \bibinfo{author}{Heinrich, B.} \&
  \bibinfo{author}{Aharoni, A.}
\newblock \bibinfo{title}{{Point singularities and magnetization reversal in
  ideally soft ferromagnetic cylinders}}.
\newblock \emph{\bibinfo{journal}{IEEE Transactions on Magnetics}}
  \textbf{\bibinfo{volume}{15}}, \bibinfo{pages}{1228--1235}
  (\bibinfo{year}{1979}).

\bibitem{AS2016}
\bibinfo{author}{Ackerman, P.~J.} \& \bibinfo{author}{Smalyukh, I.~I.}
\newblock \bibinfo{title}{{Static three-dimensional topological solitons in
  fluid chiral ferromagnets and colloids}}.
\newblock \emph{\bibinfo{journal}{Nature Materials}}
  \textbf{\bibinfo{volume}{16}}, \bibinfo{pages}{426--432}
  (\bibinfo{year}{2016}).

\bibitem{Lee1992}
\bibinfo{author}{Lee, T.} \& \bibinfo{author}{Pang, Y.}
\newblock \bibinfo{title}{Nontopological solitons}.
\newblock \emph{\bibinfo{journal}{Physics Reports}}
  \textbf{\bibinfo{volume}{221}}, \bibinfo{pages}{251 -- 350}
  (\bibinfo{year}{1992}).
\newblock
  \urlprefix\url{http://www.sciencedirect.com/science/article/pii/0370157392900647}.

\bibitem{SM1979}
\bibinfo{author}{Malozemoff, A.} \& \bibinfo{author}{Slonczewski, J.}
\newblock \bibinfo{title}{Iv - domain-wall statics}.
\newblock In \bibinfo{editor}{Malozemoff, A.} \& \bibinfo{editor}{Slonczewski,
  J.} (eds.) \emph{\bibinfo{booktitle}{Magnetic Domain Walls in Bubble
  Materials}}, \bibinfo{pages}{77 -- 121} (\bibinfo{publisher}{Academic Press},
  \bibinfo{year}{1979}).
\newblock
  \urlprefix\url{http://www.sciencedirect.com/science/article/pii/B978012002951850008X}.

\bibitem{Miltat2002}
\bibinfo{author}{Miltat, J.} \& \bibinfo{author}{Thiaville, A.}
\newblock \bibinfo{title}{Vortex cores--smaller than small}.
\newblock \emph{\bibinfo{journal}{Science}} \textbf{\bibinfo{volume}{298}},
  \bibinfo{pages}{555--555} (\bibinfo{year}{2002}).
\newblock \urlprefix\url{https://science.sciencemag.org/content/298/5593/555}.
\newblock
  \eprint{https://science.sciencemag.org/content/298/5593/555.full.pdf}.

\bibitem{Kerr99}
\bibinfo{author}{Kerr, R.~M.} \& \bibinfo{author}{Brandenburg, A.}
\newblock \bibinfo{title}{Evidence for a singularity in ideal
  magnetohydrodynamics: Implications for fast reconnection}.
\newblock \emph{\bibinfo{journal}{Phys. Rev. Lett.}}
  \textbf{\bibinfo{volume}{83}}, \bibinfo{pages}{1155--1158}
  (\bibinfo{year}{1999}).
\newblock \urlprefix\url{https://link.aps.org/doi/10.1103/PhysRevLett.83.1155}.

\bibitem{SLCT2009}
\bibinfo{author}{Smalyukh, I.~I.}, \bibinfo{author}{Lansac, Y.},
  \bibinfo{author}{Clark, N.~A.} \& \bibinfo{author}{Trivedi, R.~P.}
\newblock \bibinfo{title}{{Three-dimensional structure and multistable optical
  switching of triple-twisted particle-like excitations in anisotropic
  fluids}}.
\newblock \emph{\bibinfo{journal}{Nature Materials}}
  \textbf{\bibinfo{volume}{9}}, \bibinfo{pages}{139--145}
  (\bibinfo{year}{2009}).
\newblock \bibinfo{note}{Article}.

\bibitem{LLZ2018}
\bibinfo{author}{Liu, Y.}, \bibinfo{author}{Lake, R.~K.} \&
  \bibinfo{author}{Zang, J.}
\newblock \bibinfo{title}{Binding a hopfion in a chiral magnet nanodisk}.
\newblock \emph{\bibinfo{journal}{Phys. Rev. B}} \textbf{\bibinfo{volume}{98}},
  \bibinfo{pages}{174437} (\bibinfo{year}{2018}).
\newblock \urlprefix\url{https://link.aps.org/doi/10.1103/PhysRevB.98.174437}.

\bibitem{DHKim2019}
\bibinfo{author}{Kim, D.-H.} \emph{et~al.}
\newblock \bibinfo{title}{{Bulk Dzyaloshinskii–Moriya interaction in
  amorphous ferrimagnetic alloys}}.
\newblock \emph{\bibinfo{journal}{Nat. Mater.}} \textbf{\bibinfo{volume}{18}},
  \bibinfo{pages}{685 -- 690} (\bibinfo{year}{2019}).
\newblock \bibinfo{note}{Article}.

\bibitem{Sutcliffe2018}
\bibinfo{author}{Sutcliffe, P.}
\newblock \bibinfo{title}{Hopfions in chiral magnets}.
\newblock \emph{\bibinfo{journal}{Journal of Physics A: Mathematical and
  Theoretical}} \textbf{\bibinfo{volume}{51}}, \bibinfo{pages}{375401}
  (\bibinfo{year}{2018}).
\newblock \urlprefix\url{https://doi.org/10.1088/1751-8121/aad521}.

\bibitem{TS2018}
\bibinfo{author}{Tai, J.-S.~B.} \& \bibinfo{author}{Smalyukh, I.~I.}
\newblock \bibinfo{title}{Static hopf solitons and knotted emergent fields in
  solid-state noncentrosymmetric magnetic nanostructures}.
\newblock \emph{\bibinfo{journal}{Phys. Rev. Lett.}}
  \textbf{\bibinfo{volume}{121}}, \bibinfo{pages}{187201}
  (\bibinfo{year}{2018}).
\newblock
  \urlprefix\url{https://link.aps.org/doi/10.1103/PhysRevLett.121.187201}.

\bibitem{Chen2013}
\bibinfo{author}{Chen, B. G.-g.}, \bibinfo{author}{Ackerman, P.~J.},
  \bibinfo{author}{Alexander, G.~P.}, \bibinfo{author}{Kamien, R.~D.} \&
  \bibinfo{author}{Smalyukh, I.~I.}
\newblock \bibinfo{title}{{Generating the Hopf Fibration Experimentally in
  Nematic Liquid Crystals}}.
\newblock \emph{\bibinfo{journal}{Phys. Rev. Lett.}}
  \textbf{\bibinfo{volume}{110}}, \bibinfo{pages}{237801}
  (\bibinfo{year}{2013}).

\bibitem{donnelly20}
\bibinfo{author}{Donnelly, C.} \emph{et~al.}
\newblock \bibinfo{title}{{Time-resolved imaging of three-dimensional nanoscale
  magnetisation dynamics}}.
\newblock \emph{\bibinfo{journal}{Nature Nanotechnology}}
  \textbf{\bibinfo{volume}{15}}, \bibinfo{pages}{356--360}
  (\bibinfo{year}{2020}).
\newblock \bibinfo{note}{Article}.

\bibitem{PU1985}
\bibinfo{author}{Pokrovskii, V.~L.} \& \bibinfo{author}{Uimin, G.~V.}
\newblock \bibinfo{title}{{Dynamics of vortex pairs in a two-dimensional
  magnetic material}}.
\newblock \emph{\bibinfo{journal}{JETP Lett.}} \textbf{\bibinfo{volume}{41}},
  \bibinfo{pages}{128} (\bibinfo{year}{1985}).

\bibitem{Papanicolaou_1999}
\bibinfo{author}{Papanicolaou, N.} \& \bibinfo{author}{Spathis, P.~N.}
\newblock \bibinfo{title}{Semitopological solitons in planar ferromagnets}.
\newblock \emph{\bibinfo{journal}{Nonlinearity}} \textbf{\bibinfo{volume}{12}},
  \bibinfo{pages}{285--302} (\bibinfo{year}{1999}).
\newblock \urlprefix\url{https://doi.org/10.1088/0951-7715/12/2/008}.

\bibitem{Cooper98}
\bibinfo{author}{Cooper, N.~R.}
\newblock \bibinfo{title}{Solitary waves of planar ferromagnets and the
  breakdown of the spin-polarized quantum hall effect}.
\newblock \emph{\bibinfo{journal}{Phys. Rev. Lett.}}
  \textbf{\bibinfo{volume}{80}}, \bibinfo{pages}{4554--4557}
  (\bibinfo{year}{1998}).
\newblock \urlprefix\url{https://link.aps.org/doi/10.1103/PhysRevLett.80.4554}.

\bibitem{huang17}
\bibinfo{author}{Huang, Y.}, \bibinfo{author}{Kang, W.},
  \bibinfo{author}{Zhang, X.}, \bibinfo{author}{Zhou, Y.} \&
  \bibinfo{author}{Zhao, W.}
\newblock \bibinfo{title}{Magnetic skyrmion-based synaptic devices}.
\newblock \emph{\bibinfo{journal}{Nanotechnology}}
  \textbf{\bibinfo{volume}{28}}, \bibinfo{pages}{08LT02}
  (\bibinfo{year}{2017}).
\newblock \urlprefix\url{http://stacks.iop.org/0957-4484/28/i=8/a=08LT02}.

\bibitem{FSFHFC2017}
\bibinfo{author}{Fern{\'a}ndez-Pacheco, A.} \emph{et~al.}
\newblock \bibinfo{title}{{Three-dimensional nanomagnetism}}.
\newblock \emph{\bibinfo{journal}{Nature Communications}}
  \textbf{\bibinfo{volume}{8}}, \bibinfo{pages}{15756} (\bibinfo{year}{2017}).

\bibitem{hollerpin17}
\bibinfo{author}{Holler, M.} \emph{et~al.}
\newblock \bibinfo{title}{Omny pin—a versatile sample holder for tomographic
  measurements at room and cryogenic temperatures}.
\newblock \emph{\bibinfo{journal}{Review of Scientific Instruments}}
  \textbf{\bibinfo{volume}{88}}, \bibinfo{pages}{113701}
  (\bibinfo{year}{2017}).

\bibitem{holler17}
\bibinfo{author}{Holler, M.} \emph{et~al.}
\newblock \bibinfo{title}{{High-resolution non-destructive three-dimensional
  imaging of integrated circuits}}.
\newblock \emph{\bibinfo{journal}{Nature}} \textbf{\bibinfo{volume}{543}},
  \bibinfo{pages}{402--406} (\bibinfo{year}{2017}).

\bibitem{ptychoreview}
\bibinfo{author}{Pfeiffer, F.}
\newblock \bibinfo{title}{{X-ray Ptychography}}.
\newblock \emph{\bibinfo{journal}{Nature Photonics}}
  \textbf{\bibinfo{volume}{12}}, \bibinfo{pages}{9--17} (\bibinfo{year}{2017}).

\bibitem{rodenburg07}
\bibinfo{author}{Rodenburg, J.~M.} \emph{et~al.}
\newblock \bibinfo{title}{Hard-x-ray lensless imaging of extended objects}.
\newblock \emph{\bibinfo{journal}{Phys. Rev. Lett.}}
  \textbf{\bibinfo{volume}{98}}, \bibinfo{pages}{034801}
  (\bibinfo{year}{2007}).
\newblock
  \urlprefix\url{https://link.aps.org/doi/10.1103/PhysRevLett.98.034801}.

\bibitem{Wakonig20}
\bibinfo{author}{Wakonig, K.} \emph{et~al.}
\newblock \bibinfo{title}{{{\it PtychoShelves}, a versatile high-level
  framework for high-performance analysis of ptychographic data}}.
\newblock \emph{\bibinfo{journal}{Journal of Applied Crystallography}}
  \textbf{\bibinfo{volume}{53}}, \bibinfo{pages}{574–--586}
  (\bibinfo{year}{2020}).
\newblock \urlprefix\url{https://doi.org/10.1107/S1600576720001776}.

\bibitem{Scagnoli09}
\bibinfo{author}{Scagnoli, V.} \emph{et~al.}
\newblock \bibinfo{title}{{Linear polarization scans for resonant X-ray
  diffraction with a double-phase-plate configuration}}.
\newblock \emph{\bibinfo{journal}{Journal of Synchrotron Radiation}}
  \textbf{\bibinfo{volume}{16}}, \bibinfo{pages}{778--787}
  (\bibinfo{year}{2009}).
\newblock \urlprefix\url{https://doi.org/10.1107/S0909049509035006}.

\bibitem{Donnelly_thesis}
\bibinfo{author}{Donnelly, C.}
\newblock \emph{\bibinfo{title}{Hard X-ray Tomography of Three Dimensional
  Magnetic Structures}}.
\newblock \bibinfo{type}{{Ph. D. Thesis}}, \bibinfo{school}{ETH Zurich}
  (\bibinfo{year}{2017}).

\bibitem{vanheel05}
\bibinfo{author}{van Heel, M.} \& \bibinfo{author}{Schatz, M.}
\newblock \bibinfo{title}{{Fourier shell correlation threshold criteria}}.
\newblock \emph{\bibinfo{journal}{{JOURNAL OF STRUCTURAL BIOLOGY}}}
  \textbf{\bibinfo{volume}{{151}}}, \bibinfo{pages}{{250--262}}
  (\bibinfo{year}{{2005}}).

\bibitem{WZ1983}
\bibinfo{author}{Wilczek, F.} \& \bibinfo{author}{Zee, A.}
\newblock \bibinfo{title}{{Linking Numbers, Spin, and Statistics of Solitons}}.
\newblock \emph{\bibinfo{journal}{Phys. Rev. Lett.}}
  \textbf{\bibinfo{volume}{51}}, \bibinfo{pages}{2250--2252}
  (\bibinfo{year}{1983}).

\bibitem{G78}
\bibinfo{author}{Gross, D.~J.}
\newblock \bibinfo{title}{{Meron configurations in the two-dimensional O(3)
  $\sigma$-model}}.
\newblock \emph{\bibinfo{journal}{Nuclear Physics B}}
  \textbf{\bibinfo{volume}{132}}, \bibinfo{pages}{439--456}
  (\bibinfo{year}{1978}).

\end{thebibliography}

\end{document}